\documentclass[showpacs,twocolumn,pre,floatfix]{revtex4}
\usepackage{psfrag,epsfig,amsfonts,amssymb,amsmath,wasysym}
\usepackage{dcolumn}

\usepackage[normalem]{ulem}
\usepackage{color}

\begin{document} 
\newcommand{\steady}{\rho_{eq}}
\newcommand{\ssteady}{\rho_{eq}^{\ast}}
\newcommand{\hr}{{\cal H}}
\newcommand{\Or}{{\cal O}}
\newcommand{\gset}{{\cal G}}
\newcommand{\Tr}{\mbox{Tr}}
\newcommand{\da}{\Delta_{\! A}}
\newcommand{\db}{\Delta_{\! B}}
\newcommand{\daf}{\Delta_{\! A-B({\bf b})}}
\newcommand{\dda}{\delta \! A}
\newcommand{\deltaA}{D}
\newcommand{\sa}{\sigma_{\! A}}
\newcommand{\mic}{\mathrm{mic}}
\newcommand{\rmi}{{\rm i}}
\newcommand{\eff}{\mathrm{eff}}

\providecommand{\norm}[1]{\|#1\|}

\newcommand{\Aerg}{{A^\ast}}
\newcommand{\RR}{{\mathbb R}}
\newcommand{\AAA}{\overline{A}}
\newcommand{\Atot}{\overline{A_{2}}}
\newcommand{\Abar}{\langle A\rangle_{eq}}
\newcommand{\sea}{{\Gamma}_{\!sea}}
\newcommand{\kam}{{\Gamma}_{\!KAM}}
\newcommand{\Gammatot}{\Gamma^2}
\newcommand{\kamtot}{\bar\Gamma^{2}_{erg}}
\newcommand{\kamtotp}{\bar\Gamma^{2+}_{erg}}
\newcommand{\limt}{\lim_{T\to\infty}\int_0^T\!\frac{dt}{T}}

\title{On the Foundation of Statistical Mechanics under Experimentally Realistic Conditions: A Comparison between the Quantum and the Classical Case}
\author{Peter Reimann}
\author{Mykhaylo Evstigneev}
\affiliation{Universit\"at Bielefeld, Fakult\"at f\"ur Physik, 33615 Bielefeld, Germany}

\begin{abstract}
Focusing on isolated macroscopic systems,
described either in terms of a
quantum mechanical or a classical 
model, our two key questions are:
In how far does an initial ensemble
(usually far from equilibrium and
largely unknown in detail)
evolve towards a stationary long-time behavior 
(``equilibration'')?
In how far is this steady state in 
agreement with the microcanonical
ensemble as predicted by Statistical Mechanics
(``thermalization'')?
In the first part of the paper, a recently
developed quantum mechanical treatment of the
problem is briefly summarized,
putting particular emphasis on the 
realistic modeling of experimental measurements 
and non-equilibrium initial conditions.
Within this framework, equilibration can be
proven under very weak assumptions about
those measurements and initial conditions, 
while thermalization still requires quite strong 
additional hypotheses.
In the second part, an analogous approach within 
the framework of classical mechanics is developed
and compared with the quantum case.
In particular, the assumptions to guarantee
classical equilibration are now rather strong,
while thermalization then follows under 
relatively weak additional conditions.
\end{abstract}

\maketitle

\section{Introduction}
\label{s1}
According to textbook Statistical Physics,
all properties of a macroscopic system at 
thermal equilibrium are perfectly described 
by the canonical 
ensemble when weakly coupled to a thermal 
bath and by the microcanonical ensemble 
when isolated from the rest of the world.
However, there still seems to be no truly 
satisfactory explanation of why this 
is so \cite{skl93,pen79,gem09}.

Here we adopt the most common viewpoint
that investigations of this problem should 
be based on standard Quantum or Classical Mechanics 
and should start with the treatment of
isolated systems.
Indeed, it is widely believed that by 
considering a system in contact with a 
thermal bath as a single, isolated 
``supersystem'', the canonical 
formalism should be deducible from 
the microcanonical one.
In turn, to directly describe an open system 
alone (without the bath to which it is coupled)
by means of standard Quantum or Classical 
Mechanics does not seem possible.

While it is true that real systems
are never perfectly isolated, 
it seems a quite reasonable working hypothesis
(supported also by numerical simulations) that
modeling them as perfectly isolated
still provides a suitable description
of the actual reality \cite{skl93,wig71}.
In the same vein, though classical models
may possibly be inappropriate to describe 
{\em any} real macroscopic system \cite{wal}, 
it remains interesting to understand why classical 
Statistical Mechanics agrees so well with 
many experiments and computer simulations of 
classical many body systems.

Excellent reviews of the large number 
of pertinent ``classic'' (i.e. older) 
works and the ongoing debate about their 
physical implications is provided e.g.
by \cite{skl93,pen79,far64,jan69}.
For more recent developments we refer to 
\cite{spec,pol11,yuk11}
and references therein.
With our present work we further pursue
a recently developed new approach 
to the following two fundamental problems
\cite{rei08,lin09,lin10,rei10,sho11,sho12,rei12}.
Equilibration: 
In how far does a non-equilibrium initial 
ensemble evolve towards a stationary 
long-time behavior?
Thermalization: In how far is this steady state 
in agreement with the microcanonical ensemble
predicted by equilibrium Statistical Mechanics?

A first key point of our approach is the 
modeling concept \cite{pen79} 
that any given experimental 
system gives rise to a specific, 
well defined statistical ensemble 
(quantum mechanical density operator or
classical probability density)
the details of which are, however, 
unknown in practice.
Typically, one only knows that 
in the initial state
certain (usually macroscopic) observables are
relatively sharply distributed about some
approximately known mean values.
A main challenge of the theory is to properly
cope with this lack of knowledge.
A second key point is not to modify or 
approximate in any way the exact time 
evolution 
according to standard Quantum 
or Classical Mechanics.
The third key point is to focus on 
experimentally realistic observables,
exhibiting a finite range and a 
finite resolution.

For the rest, our present approach still 
covers essentially arbitrary (generic)
systems, and as such
is complementary to many recent 
investigations of different specific 
quantum mechanical 
model classes, observables, and initial conditions
-- often with a main focus on the so-called eigenstate
thermalization hypothesis and
the role of quantum (non-)integrability in this context
-- see e.g.\ 
\cite{tas98,sre99,caz06,rig07,kol07,man07,cra08,bar08,rig08,spec,rou10,pal10,bri10,gog11,ike11,ban11,pol11,cam11,jac11,kas11,ji11,can11,kas12,rig12}
and  references therein.
Indeed, since the ``basic laws'' of Statistical Physics are
supposed to be of largely universal validity, also the foundation 
of such laws should not focus on any specific class of models, 
observables, or initial conditions.

In the first part of our paper 
(Sects. \ref{s2}-\ref{s4}) we briefly summarize
some previously obtained results in the Quantum Mechanical 
case from Refs. \cite{rei08,lin09,lin10,rei10,sho11,sho12,rei12}.
In the subsequent main part of the paper 
(Sects. \ref{s5}-\ref{s8})
the corresponding Classical Mechanical
approach is outlined and compared with the 
Quantum Mechanical counterpart.
Our present approach differs in the 
following important respects from some
other well known approaches, especially 
within the realm of 
``Ergodic Theory'' 
\cite{skl93,far64,jan69}.

First, it is not justified to only consider
averages over sufficiently or even
infinitely long time intervals
in order to theoretically account for real 
measurements, as pointed out
particularly clearly e.g. by Sklar
(p. 176 in \cite{skl93}):
``If our macroscopic measurements could all be 
legitimately constructed as infinite time averages,
then every macroscopic measurement would have to result 
in the equilibrium value for the quantity in
question. We could have no ...  ability 
to track the approach to equilibrium by following the 
variation of these quantities as they approached 
their final equilibrium values.
But of course we can macroscopically determine the 
existence of non-equilibrium states''.
As a matter of fact, there can be little doubt 
that Statistical Mechanics is actually
supposed to apply to arbitrarily ``short'' 
measurements, in agreement with all so far 
experimental and numerical experience.
In our present approach, the main emphasis
is on ``instantaneous'' rather than on  
time-averaged measurements.

Second, it is not sufficient to
focus on macroscopic 
observables, as done e.g. 
in \cite{khi60,wig71,leb}.
Quite to the contrary, equilibrium Statistical Mechanics 
in fact also covers microscopic observables, 
e.g. the position, velocity etc. of one 
specific ``tracer'' particle within a fluid, 
and its pertinent (probabilistic)
predictions are in perfect agreement with 
numerical simulations and with 
the rapidly increasing number of 
experiments on single molecules, 
nano-particles etc.
In our present approach, arbitrary observables
(with experimentally realistic range and resolution) 
are admitted.

Third, it is not sufficient 
to focus on an arbitrary but fixed energy (hyper-)surface,
as usually done in classical ergodic theory 
\cite{far64,jan69}.
The reason is that the above mentioned statistical
ensemble which accounts for the given experimental
systems is largely unknown but certainly cannot be 
realistically assumed to exhibit an infinitely 
sharp energy distribution.
Our present approach admits a finite (and in detail 
unknown) spread of the energy distribution.

Fourth, the so-called ergodic hypothesis 
in its traditional form cannot be taken for 
granted, as done e.g. in 
\cite{far64,jan69,hob71}.
While this hypothesis has been commonly 
considered as not yet proven but 
``probably true'' for a long time,
the work by Markus and Meyer 
\cite{mar78}, KAM-theory, and numerical 
evidence are making it more and more clear 
that typical Hamiltonian systems 
exhibit a so-called divided phase space,
consisting of a chaotic (ergodic) 
component of positive measure 
(chaotic sea)
and a regular (non-ergodic) component
(union of the so-called KAM tori)
of positive measure \cite{skl93,bun08}.
In our present approach, this fact
is taken into account.

\section{Quantum Mechanical Framework}
\label{s2}
We consider a large (macroscopic but finite), 
isolated system, modeled 
in terms of a 
Hilbert space $\hr$
and a time-independent Hamiltonian $H :\hr\to\hr$ 
of the form
\begin{equation}
H = \sum_n E_n P_n \ ,
\label{1}
\end{equation}
where the $P_n$ are projectors onto the 
eigenspaces of $H$ with mutually different
eigenvalues $E_n$ and multiplicities
\begin{equation}
\mu_n:=\Tr \{ P_n \} \ .
\label{5}
\end{equation}

System states are described by
density operators $\rho (t)$, 
evolving according to
$\rho(t)=U_t\rho(0) U_t^\dagger$
with propagator
$U_t:=\exp\{-\rmi Ht\}$ and $\hbar =1$.
Exploiting (\ref{1}) we thus can conclude that
\begin{equation}
\rho(t)=\sum_{mn} \rho_{mn}(0) \exp[-\rmi (E_m-E_n)t] \ ,
\label{6}
\end{equation}
\begin{equation}
\rho_{mn}(t):=P_m\rho(t)P_n \ .
\label{7}
\end{equation}
While we will be mainly interested in density 
operators describing some statistical ensemble 
(i.e. mixed states $\rho$ with $\Tr\{\rho^2\}<1$), 
it is nevertheless worth to point out that 
formally our considerations will also
cover pure states ($\Tr\{\rho^2\}=1$) 
as special case.

Observables are represented by self-adjoint
operators $A$ with expectation values
$\Tr\{\rho(t)A\}$.
In order to model real experimental 
measurements it is, however,
not necessary to admit any arbitrary self-adjoint 
operator \cite{khi60,realobs2,realobs3,realobs4,realobs5,lof,geo95,pop06}.
Rather, it is sufficient to focus on
experimentally realistic observables 
in the following sense \cite{rei08,rei08a,rei10}: 
Any observable $A$ must represent an experimental 
device with a finite range of possible 
outcomes of a measurement,
\begin{equation}
\da := 
\sup_{\norm{\psi}=1} \langle\psi|A|\psi\rangle
- \inf_{\norm{\psi}=1} \langle\psi|A|\psi\rangle \ ,
\label{8}
\end{equation}
where the supremum and infimum are taken over all 
normalized vectors $|\psi\rangle\in \hr$.
Moreover, this working range $\da$
of the device must be limited to experimentally 
reasonable values compared to its 
resolution limit $\dda$.
All measurements known to the present author 
yield less than 20 relevant digits, i.e.\ 
\begin{equation}
\da/\dda \leq 10^{20} \ .
\label{9}
\end{equation}
Maybe some day 100 or 1000 relevant digits will
become feasible, but it seems reasonable that a theory
which does not go very much beyond that will do.
Note that similar restrictions also apply
to ``numerical experiments'' by 
computer simulations.

The specific observable $A=P_n$ describes
the population of the (possibly degenerate)
energy level $E_n$ with expectation value 
(occupation probability)
\begin{equation}
p_n:=\Tr\{ P_n \rho(t) \} .
\label{11}
\end{equation}
Note that $P_n$ commutes with $H$ from (\ref{1})
and hence the level populations $p_n$ are 
conserved quantities.

For a system with $f$ degrees of freedom
there are roughly $10^{\Or(f)}$ energy eigenstates 
with eigenvalues in every interval of $1$J beyond the ground 
state energy, see e.g.  \cite{lldiu} or section 2.1 in \cite{rei10}.
The same estimate carries over to the number
of energy eigenvalues under the assumption that
their multiplicities (\ref{5})
are much smaller than $10^{\Or(f)}$.
For a macroscopic system with
$f=\Or(10^{23})$, the energy levels are thus 
unimaginably dense on any decent energy 
scale and even the most careful experimentalist 
will not be able to populate only a few of them
with significant probabilities $p_n$,
in particular after averaging over 
the ensemble, i.e. over many 
repetitions of the experiment.
In the generic case we thus expect
\cite{rei08,rei08a,rei10}
that -- even if the system's energy is fixed up 
to an extremely small experimental uncertainty,
and even if the energy levels are populated 
extremely unequally --
the largest ensemble-averaged 
level population $p_n$ will
be extremely small (compared to $\sum_n p_n=1$)
and typically satisfy the 
rough estimate
\begin{equation}
\max_n\, p_n=10^{-\Or(f)} \ .
\label{12}
\end{equation}

\section{Equilibration in the Quantum Case}
\label{s3}
Generically, the density operator
$\rho(t)$ is not stationary right from 
the beginning, in particular for an initial
condition $\rho(0)$ out of equilibrium.
But if the right hand side of (\ref{6})
depends on $t$ initially, it cannot 
approach for large $t$ any time-independent 
``equilibrium ensemble'' whatsoever.
In fact, any mixed state $\rho(t)$ 
returns arbitrarily close (with respect to some 
suitable distance measure in Hilbert space) 
to its initial state $\rho(0)$ for certain, 
sufficiently large times $t$, 
as demonstrated for instance in 
appendix D of reference \cite{hob71}.
We will therefore focus on the
weaker notion of equilibration
from \cite{rei08,lin09,lin10,rei10,sho11,sho12,rei12},
which merely requires the
existence of a 
time-independent ``equilibrium state'' 
$\rho_{eq}$ (density operator) with the property 
that the difference
\begin{equation}
\sigma(t):=\Tr\{\rho(t) A\}-\Tr\{\steady A\}
\label{17}
\end{equation}
between the true expectation value 
$\Tr\{\rho(t) A\}$ and the 
equilibrium reference
value $\Tr\{\steady A\}$ is unresolvably 
small for the overwhelming majority 
of times $t$ contained in any sufficiently
large (but finite) time interval $[0,T]$.
In other words, the expectation values 
$\Tr\{\rho(t) A\}$ may still exhibit 
everlasting small fluctuations around their 
``equilibrium value'' $\Tr\{\steady A\}$, 
as well as very rare
large excursions away from equilibrium 
(including the above mentioned recurrences \cite{pen79}), 
but quantitatively these 
fluctuations are either unobservably small
compared to any reasonably achievable 
resolution limit, or exceedingly rare
on any realistic time scale
after initial transients have died out.
(Note that those initial transients become 
irrelevant if the time interval $[0,T]$ 
is chosen large enough.)

It seems quite plausible that if there is any such 
equilibrium ensemble
$\steady$ at all, then it should be given 
by the time-independent part of $\rho(t)$
from (\ref{6}), i.e. 
\begin{equation}
\steady:=\sum_n\rho_{nn} \ .
\label{18}
\end{equation}
Note that the time-arguments of $\rho_{nn}$ 
have been omitted since these are conserved 
quantities according to (\ref{6}), (\ref{7}).

Technically speaking (the reasons will become 
obvious below) the quantity of foremost 
interest is the time-averaged variance
\begin{equation}
\overline{ \sigma^2(t)}^T
:=\frac{1}{T}\int_0^T dt \, \sigma^2(t) 
\label{n1}
\end{equation}
following from (\ref{17}) and (\ref{18}).
Considering and estimating such averages
has a long tradition, see e.g. 
\cite{lud58,boc59,far64,jan69,per84,deu91,sre96,tas98}.
Substantial progress along this line
has been achieved quite recently in the works 
\cite{rei08,lin09,lin10,rei10,sho11,sho12,rei12}.
A particularly strong and general result
has been obtained in 2012 by Short and 
Farrelly \cite{sho12}
(for details see also \cite{sho11,rei12}),
showing that
\begin{equation}
\overline{ \sigma^2(t)}^T
\leq \frac{1}{2} \, \da^2\, g\, \sum_n p_n^2
\label{n2}
\end{equation}
for all sufficiently large $T$. 
Here, $g$ denotes the maximal degeneracy of
energy gaps,
\begin{equation}
g := 
\max_{m\not = n}|\{(k,l) \, : \, E_k-E_l=E_m-E_n\}| \ ,
\label{67}
\end{equation}
where $|S|$ stands for the number of elements
contained in the 
set $S$.
In other words, $g$ is the maximal number of (exactly) 
coinciding energy differences among all possible pairs of 
distinct energy eigenvalues
of the Hamiltonian $H$ from (\ref{1}).
In particular, Hamiltonians with
degenerate energy eigenvalues and 
degenerate energy gaps are thus 
still admitted in (\ref{n2}).
We further remark that the upper 
limit by Short and Farrelly 
\cite{sho12} for the minimal 
$T$-value admitted in (\ref{n2})
depends on the spectrum 
of $H$ and on the level populations $p_n$, 
but not on the observable $A$.
Finally, we note that the quite obvious relations
\begin{eqnarray}
{\sum_n} p_n^2 \leq {\max_n\,} p_n
{\sum_n} p_n = {\max_n\,} p_n,
\label{14}
\\
{\max_n\,} p_n = \big( {\max_n\,} p_n^2 \big)^{1/2} 
\leq \big( {\sum_n} p_n^2 \big)^{1/2},
\label{15}
\end{eqnarray}
readily lead to the conclusion
\begin{equation}
{\max_n\,} p_n \; \mbox{small}\ \Leftrightarrow\ 
{\sum_n} p_n^2\; \mbox{small} \ .
\label{16}
\end{equation}
Hence, focusing on large ``effective dimensions'' 
(of the state $\rho(t)$) $d_{\eff}:=1/\sum_n p_n^2$ 
as done e.g. in \cite{lin09,lin10,sho11,sho12} 
is equivalent to our present assumption of
small level populations, cf. (\ref{12}).

Next we define for any given $\dda>0$ and $T>0$
the quantity
\begin{equation}
T_{\dda} := 
\big| \{0< t  <T\, : \, |\Tr\{\rho(t)A\}-\Tr\{\steady A\}|\geq\dda\}\big| \ ,
\label{69}
\end{equation}
where $|M|$ denotes the size (Lebesgue measure) of the set $M$.
According to (\ref{17}), $T_{\dda}$ thus
represents the measure of all times $t\in [0, T]$ for which 
$|\sigma(t)|\geq \dda$ holds true.
It follows that $\sigma^2(t)\geq \dda^2$
for a set of times $t$ of measure $T_{\dda}$
and that
$\sigma^2(t)\geq 0$ for all remaining times $t$
in $[0, T]$.
Hence the temporal average (\ref{n1}) must be at least
$\dda^2 T_{\dda}/T$ and 
we can conclude from (\ref{n2}) and (\ref{14}) that
\begin{equation}
\frac{T_{\dda}}{T}\leq  \frac{1}{2}\,\left(\frac{\da}{\dda}\right)^2\, g\, 
{\max_n\,} p_n 
\label{71}
\end{equation}
for all sufficiently large $T$.

We remark that the original derivation of (\ref{n2})
from \cite{sho12} and thus the result (\ref{71})
is restricted to finite dimensional Hilbert 
spaces $\hr$.
The generalization to infinite-dimensional 
systems has been worked out in \cite{rei12},
yielding instead of (\ref{71}) the slightly 
weaker bound
\begin{equation}
\frac{T_{\dda}}{T}\leq  2\,
\left(\frac{\da}{\dda}\right)^2 \, g\,  {\max_n\,} p_n \ .
\label{74}
\end{equation}

According to (\ref{69}), the left hand side 
of (\ref{74}) represents the fraction of all times
$t\in[0, T]$ for which there is an experimentally
resolvable difference between the true expectation 
value $\Tr\{\rho(t)A\}$ and the time-independent 
equilibrium expectation value $\Tr\{\steady A\}$.
On the right hand side of (\ref{74}), $\da/\dda$ is 
the range-to-resolution ratio,
which can be considered as bounded according 
to (\ref{9}) for all experimentally realistic 
measurements $A$.
The next factor, $g$, is the maximal degeneracy 
of energy gaps from (\ref{67}).
Finally, ${\max_n\,} p_n$ represents the largest, 
ensemble-averaged occupation probability of 
the (possibly degenerate) energy 
eigenvalues $E_n$, see (\ref{11}).
Typically, one expects that the rough upper
bound (\ref{12}) applies, except if certain
energy eigenvalues
are so extremely highly degenerate that
the multiplicities defined in (\ref{5}) severely
reduce the pertinent energy level density
compared to the non-degenerate case, 
see above (\ref{12}).

For a system with sufficiently many degrees of 
freedom $f$ and no exceedingly large 
degeneracies of energy eigenvalues and energy 
gaps \cite{f1}
we thus can conclude from (\ref{74}) with 
(\ref{9}) and (\ref{12}) 
that the system behaves in every possible experimental
measurement exactly as if it were in the 
equilibrium
state $\steady$ for the overwhelming majority of times
within any sufficiently large (but finite) 
time interval $[0, T]$ \cite{f1a},
i.e. we recover ``equilibration'' in the sense 
proposed at the beginning of this section.

\section{Thermalization in the Quantum Case}
\label{s4}
Next we address the somewhat related but still 
notably different issue of ``thermalization'', i.e.,
the question whether, and to what extend, 
the above discussed equilibrium expectation 
value $\Tr\{\steady A\}$
is in agreement with that predicted by the
microcanonical ensemble.

For the sake of simplicity, we confine ourselves
to non-degenerate Hamiltonians throughout this section,
i.e., all projectors $P_n$ in (\ref{1})
are of the form $P_n=|n\rangle\langle n|$, where $|n\rangle$
denotes the energy eigenvector corresponding to the
(non-degenerate) energy eigenvalue $E_n$.
Furthermore, we assume without loss of generality that 
the eigenstates $|n\rangle$ are ordered according to 
their eigenvalues, i.e. $E_{n}\leq E_{n+1}$ for all $n$
\footnote{Note that ``gaps'' are a quite common
feature of the energy spectra of {\em single}
(quasi-) particles. In contrast, our present 
$E_n$ and $|n\rangle$ refer to the {\em many particle}
eigenenergies and eigenstates of the total (many particle)
Hamiltonian $H$ from Eq. (\ref{1}). Even if the
single-particle spectra exhibit gaps, the 
many-particle spectra are generically gapless.}.
According to (\ref{7}), (\ref{11}), (\ref{18}) it 
follows that 
\begin{equation}
\Tr\{\steady A\} =\sum_n p_n \, \langle n|A |n\rangle \ ,
\label{101}
\end{equation}
while the expectation value predicted by the 
microcanonical ensemble takes the form
\begin{equation}
\Tr\{\rho_{mic} A\} =\sum_n p^{\mic}_{n} \langle n|A |n\rangle \ ,
\label{102}
\end{equation}
where the level populations $p^{\mic}_{n}$
are equal to a normalization constant if
$E_n$ is contained within some small 
energy interval 
\begin{equation}
I:=[E_{mic}-\Delta E, \,E_{mic}] 
\label{103}
\end{equation}
and zero otherwise \cite{lldiu}.

If (\ref{101}) and 
(\ref{102}) yield measurable 
differences for experimentally 
realistic $\rho(0)$ and $A$, the 
``purely Quantum Mechanical'' prediction (\ref{101})
is commonly considered as ``more fundamental''
\cite{rig07,rig08,cra08}.
In other words, provided equilibrium Statistical 
Mechanics itself is valid at all, its prediction
(\ref{102}) must agree with (\ref{101}), 
at least within experimentally realistic resolution 
limits.

What are these validity conditions, beyond
which the microcanonical formalism of
equilibrium Statistical Mechanics may 
break down?

A first well known validity condition for 
the microcanonical formalism
is, as said below (\ref{102}), that only
$E_n$ within some small energy interval 
(\ref{103}) have a non-vanishing occupation 
probability.
More generally, in equilibrium Statistical
Mechanics it is taken for granted that the
system energy is ``fixed'' up to unavoidable
experimental uncertainties.
On the other hand,
realistic initial conditions as discussed above Eq. (\ref{12})
require that this energy uncertainty is much larger than
$10^{-\Or(f)}$ Joule, which is obviously
always fulfilled in practice, but we never
introduced or exploited any type of 
upper limit for this uncertainty so far, i.e.
the energy uncertainty may still be 
arbitrarily large in (\ref{74}) and (\ref{101}).
In other words, for large energy uncertainties,
our key relation (\ref{101}) remains valid, 
while equilibrium Statistical Mechanics is 
likely to become invalid.
This is clearly a not at all surprising case
of disagreement between (\ref{101}) 
and (\ref{102}) .

To avoid such ``almost trivial'' cases, 
we henceforth take for granted that
the system energy is known up to an uncertainty 
$\Delta E$ which is as small as possible, 
but still experimentally realistic.

A second (often tacit) validity condition 
of the microcanonical formalism is that
the expectation values (\ref{102})
are required/assumed to be (practically) 
independent of the exact choice of the 
interval $I$ in (\ref{103}), 
i.e. of its upper limit $E_{mic}$ and its
width $\Delta E$.
But essentially this means that the value 
of the sum in Eq. (\ref{101}) must be 
largely independent of the details of 
the weights $p_{n}$.
The same conclusion also follows
from the equivalence of the microcanonical 
and canonical ensembles (for all 
energies $E_{mic}$), 
considered as a self-consistency condition 
for equilibrium Statistical Mechanics 
\cite{lof,geo95}.

If the sum in Eq. (\ref{101}) has the above 
mentioned property of being largely independent 
of the weights $p_n$, we henceforth abbreviate 
this fact by saying that it has the property (P).
(A more precise definition of ``largely independent'' 
is difficult: essentially, the value of the
sum in (\ref{101}) is supposed to exhibit negligible
variations for all sets of weights $\{p_n\}$ 
arising under ``physically relevant'' circumstances.)

Given property (P) holds, the expectation values 
(\ref{101}) and (\ref{102}) are indeed practically 
indistinguishable.

Our first remark regarding property (P)
itself is that no experimentalist can control 
the populations $p_n$ of the unimaginably dense 
energy levels $E_n$, apart from the very gross
fact that they are ``mainly concentrated within the
interval $I$ from (\ref{103})''.
So, if the details of the $p_n$ in (\ref{101})
{\em would} matter, then not only 
equilibrium Statistical Mechanics would 
break down, but also reproducing measurements,
in particular in different labs, would 
be largely impossible.

The simplest way to guarantee property (P)
seems to require/assume 
that the expectation values $\langle n|A |n\rangle$
hardly vary within any small energy interval 
of the form (\ref{103}).
This is indeed part of a common 
conjecture about the semiclassical 
behavior of fully chaotic classical 
systems, see \cite{sre96,fei86} and references therein.
In particular, negligible variations of 
$\langle n|A |n\rangle$ for close by $n$-values
imply Srednicki's ``eigenstate
thermalization hypothesis'' \cite{sre94} 
(anticipated in \cite{jen85} and
revisited in Ref. \cite{rig08}), implying 
that each individual energy eigenstate 
$|n\rangle$ behaves like the equilibrium 
ensemble.

An alternative way to guarantee property (P)
follows from the argument by Peres \cite{per84}
that even if the $\langle n|A |n\rangle$ 
may notably vary with $n$,
the immense number of relevant summands in 
(\ref{101}) may
-- for ``typical'' $A$ and $\rho(0)$ -- 
lead to a kind of statistical 
averaging effect and thus a largely 
$\rho(0)$-independent overall value 
of the sum.

Numerically, the validity and possible failure
of such conjectures and of property
(P) itself have been exemplified e.g. in 
\cite{rig07,kol07,man07,rig08,pal10,bri10,gog11,ike11,ban11,cam11,jac11,ji11,can11,rig12,fei84,jen85}.
While the details -- in particular the 
role of  ``more basic'' system properties 
like ``quantum ergodicity'' and ``(non-)integrability'' --
are still not very well understood 
\cite{bar08,rig08,rou10,far64,jan69,per84,sre96,fei86,wei92},
``equilibration'' in agreement with 
(\ref{101}) was seen numerically 
in all known cases.

\section{Classical Mechanical Framework}
\label{s5}
As in Sect. \ref{s2}, we consider an isolated system
with $f\gg 1$ degrees of freedom, but now
modeled classically in terms a 2f-dimensional
phase space $\Gamma$ and a 
time-independent Hamiltonian $H :\Gamma\to\RR$.
Given an initial condition $\phi_0\in\Gamma$ at
time $t_0=0$,
the state at time $t$ can be written in terms of the
propagator $\gamma_t$ induced by $H$
as $\phi(t)=\gamma_{t}(\phi_0)$.
The fact that in reality the actual 
microstate $\phi(t)$ is not exactly 
known is modeled by means of a probability 
density $\rho(\phi,t)$, being non-negative, 
Lebesque integrable, and
normalized on $\Gamma$ \cite{pen79}. 
Further, $\rho(\phi,t)$  evolves 
from the initial ensemble $\rho_0(\phi):=\rho(\phi,0)$ according to
\begin{equation}
\rho(\phi,t)=\rho_0(\gamma_{-t}(\phi)) \ .
\label{0}
\end{equation}
In particular, pure states of the form 
$\rho(\phi,t)=\delta(\phi-\gamma_t(\phi_0))$
are thus excluded. The main place, where this
assumption turns out to be indispensable 
will be in Eq. (\ref{18j1}) below.

The resulting expectation value for an observable 
$A(\phi)$ is given by
\begin{equation}
\langle A\rangle_t:= 
\int  d\phi\, \rho(\phi,t)\, A(\phi) \ ,
\label{40}
\end{equation}
where the integral extends over the entire phase 
space $\Gamma$ and where we tacitly focus on 
Lebesgue integrable $A(\phi)$.

From Liouville's Theorem and
$H(\gamma_t(\phi))=H(\phi)$ (energy conservation)
we can infer that
\begin{equation}
\int d\phi\, a(\phi)\, b(H(\phi))
=\int d\phi\, a(\gamma_t(\phi))\, b(H(\phi))
\label{3}
\end{equation}
for arbitrary functions $a\, :\, \Gamma\to\RR$, 
$b\, :\, \RR\to\RR$, and times $t\in\RR$.
With (\ref{0}) we thus can rewrite (\ref{40}) as
\begin{equation}
\langle A\rangle_t:= 
\int  d\phi \, \rho_0(\phi) \, A(\gamma_t(\phi)) \ .
\label{40a}
\end{equation}

As in (\ref{8}), we focus on 
observables $A(\phi)$ with a finite range
\begin{equation}
\da := \sup_{\phi} A(\phi)-\inf_{\phi} A(\phi) \ ,
\label{53}
\end{equation}
where the supremum and infimum are taken over all
$\phi\in\Gamma$.
Moreover, the ratio between this range and 
the resolution limit $\dda$ is again
assumed to satisfy (\ref{9}).

Before turning to the discussion of the admissible
initial conditions $\rho_0(\phi)$, 
in particular the classical
counterpart of (\ref{12}), some very important but
somewhat technical 
preliminaries are required.
We start by recalling Birkhoff's Theorem, 
asserting \cite{far64,jan69} that
for any (Lebesgue integrable) $A(\phi)$ and
almost all $\phi\in\Gamma$
the time average
\begin{equation}
\AAA(\phi):=\limt \, A(\gamma_{t}(\phi))
\label{1b}
\end{equation}
exists and is a constant of motion, i.e.
$\AAA (\gamma_t (\phi))= \AAA (\phi)$ for all $t$.
Furthermore, based on numerical evidence, 
heuristic reasoning, and rigorous arguments in special cases,
it is commonly conjectured
\cite{skl93,for92,bun08,wig71,lic83}
that a generic Hamiltonian system
(with $f>1$) exhibits 
a so-called divided phase space.
More precisely, the total phase space $\Gamma$ 
can be decomposed into a set
$\kam\subset\Gamma$, being the 
union of all the so-called 
Kolmogorov-Arnold-Moser (KAM) tori, 
and a complement 
\begin{equation}
\sea:=\Gamma\setminus\kam \ ,
\label{1c}
\end{equation}
essentially consisting of the so-called 
``chaotic sea'' (or ``stochastic region''),
but for the sake of convenience will henceforth
be understood to also contain
a ``non-chaotic set'' of measure zero 
(e.g. so-called hyperbolic fixed points 
and periodic orbits).
Their main properties are as follows
\cite{skl93,for92,bun08,wig71,lic83}:

(i) The dynamics leaves $\sea$ 
invariant, i.e. if $\phi\in\sea $ 
then $\gamma_t(\phi)\in\sea $ for 
all $t$.
As a consequence, a similar invariance property applies to 
$\kam$.

(ii) The dynamics has the following so-called 
ergodicity property \cite{har40,hal49}:
For almost all phase points $\phi \in \sea $ 
which furthermore belong to an arbitrary but
fixed energy surface (i.e. $H(\phi)=E$ with
an arbitrary but fixed $E$), the time-average
in (\ref{1b}) is given by one and the same value.
This value will however in general be different
for different energies $E$.
In other words, for any $A(\phi)$ there exists a 
function $\Aerg:\RR\to\RR$ with the property that
\begin{equation}
\AAA(\phi)=\Aerg(H(\phi))\ \ \mbox{for almost all $\phi\in\sea $} \ .
\label{30}
\end{equation}

(iii) The dynamics has the following 
so-called (strong) mixing property \cite{har40,hal49}:
Almost all phase points $\phi,\phi' \in \sea $ 
with $H(\phi)=H(\phi')$ (i.e. $\phi$ and
$\phi'$ belong to the same energy surface)
exhibit an independent evolution over long times
in the following 
sense \cite{f2}:
\begin{eqnarray}
& & \limt \, A(\gamma_{t}(\phi))\, A(\gamma_{t}(\phi'))=
\AAA(\phi)\, \AAA(\phi')
\nonumber
\\
& & 
\mbox{for almost all $\phi,\phi'\in\sea $ with $H(\phi)=H(\phi')$} 
\label{1a}
\end{eqnarray}
In particular, if $\phi'$ happens to coincides with $\phi$
then the long time evolution of $\phi'$ is obviously not 
any more independent from that of $\phi$ and thus the 
above relation generically breaks down. 

(iv) The set $\kam$ of all KAM-tori
exhibits an extremely convoluted, nowhere 
dense, self-similar structure.
For instance, let us focus on phase points up to some
arbitrary but fixed energy $E$ and accordingly define
the restricted sets
$\Gamma(E):=\{\phi\in\Gamma \, : \, H(\phi)\leq E\}$,
$\kam(E):=\Gamma(E)\cap\kam$,
and
$\sea(E):=\Gamma(E)\cap\sea$.
Then, for all sufficiently large but finite $E$,
all three sets $\Gamma(E)$, $\kam(E)$, and
$\sea(E)$ are of positive but finite measure
(see e.g. Ref. \cite{bun08} and page 175 of Ref. \cite{skl93}).
Moreover, it is commonly expected 
\cite{skl93,bun08,lic83,alt07,bal71}
that for systems with sufficiently 
many degrees of freedom $f$
the ratios $\kam(E)/\Gamma(E)$
become arbitrarily small.
Accordingly, it seems plausible that even the most careful
experimentalist will not be able to populate 
the set $\kam$ of all KAM-tori
with significant probability
on the average over many repetitions of the 
experiment.
In other words, we expect that the initial ensemble 
$\rho_0(\phi)$ has the property that
\begin{eqnarray}
\int_{\kam} d\phi \ \rho_0(\phi)\ \ 
& &  \mbox{becomes arbitrarily small}
\nonumber
\\[-0.3cm]
& & \mbox{for sufficiently large $f$,}
\label{31}
\end{eqnarray}
where $\int_{\kam}$ indicates an integration 
over the subset $\kam$ of the total phase space 
$\Gamma$, and the word ``small'' is meant in 
comparison with $\int d\phi \ \rho_0(\phi)=1$.

This property (\ref{31}) represents the announced
classical counterpart of (\ref{12}).
In fact, according to the most common conjecture 
\cite{lic83,alt07}, an exponential decay with
$f$ analogous to (\ref{12}) may be expected also
in (\ref{31}). Explicit estimates are provided
e.g. at the end of Sec. 6.5a in Ref. \cite{lic83}
and in the paper by Falcioni et al. cited in 
Ref. \cite{alt07},  suggesting an exponential 
decay analogous to (\ref{12}).

We remark that it is always possible to
find sufficiently small hyper-spheres (or hyper-cubes etc.)
$\Gamma'\subset\Gamma$ 
with the property that the measure 
of $\kam\cap\Gamma'$ approaches that of 
$\Gamma'$ arbitrarily closely \cite{wig71}.
Condition (\ref{31}) expresses the 
expectation that the experimentalist cannot
``hit'' those tiny subsets with any notable 
probability, analogous to the unlikeliness of
``hitting'' any single energy eigenvalue 
expressed by (\ref{12}), and in spite of
the fact that initial ensembles 
$\rho_0(\phi)$ far from equilibrium
are usually concentrated themselves 
within very small phase space regions.
On the other hand, not ``hitting'' the
subset $\kam$ at all, so that the integral (\ref{31})
becomes strictly zero, also seems practically
impossible.

\section{Equilibration in the Classical Case}
\label{s6}
Proceeding like in Sect. \ref{s3}, the quantity
of foremost interest is the deviation
\begin{equation}
\sigma(t):=\langle A\rangle_t-\langle A\rangle_{eq}
\label{17a}
\end{equation}
of the true expectation value (cf. (\ref{40}))
from the time-independent expectation 
value \cite{f3}
\begin{equation}
\langle A\rangle_{eq} :=
\int d\phi\, \steady(\phi)\,  A(\phi)
\label{17b}
\end{equation}
for a suitably defined 
equilibrium
ensemble $\steady(\phi)$.
As one might have expected, the proper
choice of the latter ensemble turns out to 
be the time-averaged ``true'' ensemble 
$\rho(\phi,t)$, i.e.
\begin{equation}
\steady(\phi) :=\limt \, \rho(\phi,t) \ .
\label{18a}
\end{equation}
According to (\ref{0}) and Birkhoff's Theorem (\ref{1b}),
the limit in (\ref{18a}) indeed exists 
(for almost all $\phi$; the remaining 
exceptional $\phi$ are irrelevant in 
(\ref{17b})).

Exploiting (\ref{18a}), 
the equilibrium expectation value from (\ref{17b})
can be rewritten by means of 
(\ref{40}), (\ref{40a}), (\ref{1b}) as
\begin{equation}
\langle A\rangle_{eq}
=
\limt \langle A\rangle_t= \int d\phi 
\, \rho_0(\phi)\, \AAA(\phi) \ .
\label{18b}
\end{equation}
We further can conclude from (\ref{40a}),
(\ref{17a}), and (\ref{18b}) that
\begin{eqnarray}
\sigma(t)
& = & 
\int d\phi \, \rho_0(\phi) [A(\gamma_t(\phi))-\AAA(\phi)]
\nonumber
\\
& = & \int dE \, p(E) \deltaA (E,t)
\label{18c}
\\
\deltaA (E,t) & := & \int d\phi\, \rho (E,\phi)
[A(\gamma_t(\phi))-\AAA(\phi)]
\label{18d}
\\
p(E) & := & \int d\phi\, \delta(H(\phi)-E)\, \rho_0(\phi)
\label{18d2}
\\
\rho(E,\phi) & := & \frac{\delta(H(\phi)-E)\, 
\rho_0(\phi)}{p(E)} \ ,
\label{18d1}
\end{eqnarray}
where the integral
in (\ref{18c}) and the 
definition (\ref{18d1}) are tacitly restricted to 
$E$-values with $p(E)>0$ \cite{f4}.
Noting that $p(E)$ from (\ref{18d2}) is non-negative we 
can infer from (\ref{18c}) that
\begin{eqnarray}
\sigma^2(t)
& \leq &  \left[\int dE \, p(E) |\deltaA (E,t)|\right]^2 
\nonumber
\\
& = &
\left[\int dE \, \sqrt{p(E)}\, \sqrt{q(E)}\right]^2
\label{18g}
\\
q(E) & := & p(E)\, \deltaA^2(E,t)\ .
\label{18h}
\end{eqnarray}
Applying the Cauchy-Schwarz inequality it follows that
\begin{eqnarray}
\sigma^2(t)
& \leq &
\int dE' \, p(E')\, \int dE \, q(E) \ .
\label{18h1}
\end{eqnarray}
According to (\ref{18d2}) the first integral is unity.
With (\ref{18h}) and by indicating time averages by an 
overbar as in (\ref{n1}), it follows that
\begin{eqnarray}
\overline{\sigma^2(t)}^T
& \leq & 
\int dE \, p(E)\, \overline{\deltaA^2(E,t)}^T \ .
\label{18i}
\end{eqnarray}
In view of (\ref{18d}) the last factor takes the form
\begin{equation}
\overline{\deltaA^2(E,t)}^T
=
\int d\phi\int d\phi'\, \rho (E,\phi)\,\rho (E,\phi') \, C(T,\phi,\phi') \ ,
\label{18j}
\end{equation}
where we have introduced 
\begin{eqnarray}
C(T,\phi,\phi') & := & \overline{[A(\gamma_t(\phi))-\AAA(\phi)]\, [A(\gamma_t(\phi'))-\AAA(\phi')]}^T 
\nonumber
\\
& = & \overline{A(\gamma_t(\phi))\, A(\gamma_t(\phi'))}^T
- \overline{A(\gamma_t(\phi))}^T \, \AAA(\phi')
\nonumber
\\
& - & \AAA(\phi)\, \overline{A(\gamma_t(\phi'))}^T 
+\AAA(\phi)\, \AAA(\phi') \ .
\label{18k}
\end{eqnarray}
In the last two lines, the second and third terms have well defined
limits $T\to\infty$ for almost all $\phi$, $\phi'$
according to Birkhoff's Theorem, see (\ref{1b}).
To establish that the same applies for the first term,
we consider an auxiliary dynamics on the product space 
$\Gammatot := \Gamma \times\Gamma$, governed by the 
Hamiltonian $H_{tot}(\phi_{tot}):=H(\phi)+H(\phi')$
with $\phi_{tot}:=(\phi,\phi')\in\Gammatot$.
Since this gives rise to a well-defined Hamiltonian dynamics,
the existence of the limit in question follows by
applying Birkhoff's Theorem to this dynamics.
Under the tacit assumption that the limit $T\to\infty$
commutes with the integrations in (\ref{18i}) 
and (\ref{18j}), the right hand side in (\ref{18i}) 
thus exhibits a well defined limit $T\to\infty$ 
\cite{f5}. Denoting this limit by 
$\overline{\sigma^2(t)}^\infty$, 
it follows for any given $\epsilon>0$ that $|\overline{\sigma^2(t)}^T-\overline{\sigma^2(t)}^\infty|\leq\epsilon$ 
for all sufficiently large $T$.
Hence,
\begin{eqnarray}
\overline{\sigma^2(t)}^T
& \leq & \epsilon +
\int dE \, p(E)\, \overline{\deltaA^2(E,t)}^\infty
\label{18i1}
\end{eqnarray}
for all sufficiently large $T$
\footnote{Since $\overline{\sigma^2(t)}^T$ does not necessarily 
approach $\overline{\sigma^2(t)}^\infty$ from below,
$\epsilon$ is not superfluous in (\ref{18i1}).}.

Next, the integration domain $\Gamma\times\Gamma$ 
of the double integral in (\ref{18j}) is decomposed 
by exploiting that $\Gamma=\sea\cup\kam$
(see (\ref{1c})) into the four subdomains
$\sea\times\sea$,
$\sea\times\kam$,
$\kam\times\sea$,
and
$\kam\times\kam$.
For the sake of later convenience,
we re-unify the second and the 
fourth of them into $\Gamma\times\kam$.
Hence, we are left with the
three subdomains $\sea\times\sea$, 
$\Gamma\times\kam$, and $\kam\times\sea$.
Noting that in (\ref{18j}) we have 
$H(\phi)=H(\phi')=E$ according to (\ref{18d1}), 
it follows with (\ref{1b}), (\ref{1a}), and (\ref{18k}) 
that the contribution from the first subdomain 
$\sea\times\sea$ to (\ref{18j}) approaches 
zero in the limit $T\to\infty$,
\begin{equation}
\lim_{T\to\infty}
\int_{\sea} \!\!\!\!\! d\phi\int_{\sea} \!\!\!\!\!  d\phi'\, 
\rho (E,\phi)\,\rho (E,\phi') \, C(T,\phi,\phi')
= 0 \ .
\label{18j1}
\end{equation}
We remark that in the case of a pure state of the form
$\rho(\phi,t)=\delta(\phi-\gamma_t(\phi_0))$ 
the right hand side of (\ref{18j}) becomes equal to 
$C(T,\phi_0,\phi_0)$, approaching 
$\overline{A^2(\gamma_t(\phi_0))}^\infty-
[\AAA(\phi_0)]^2$ for $T\to\infty$
according to (\ref{18k}).
As pointed out below (\ref{1a}), this difference is
non-zero for any generic $\phi_0\in \sea$
and hence the same applies to the left hand side
of (\ref{18j1}).

Next, denoting by $K$ the modulus of the contribution 
to (\ref{18j}) from the second subdomain $\Gamma\times\kam$ 
(see above (\ref{18j1})) it follows that
\begin{equation} 
K\leq \int_{\Gamma} d\phi\int_{\kam} d\phi'\, \rho (E,\phi)\,\rho (E,\phi') \, |C(T,\phi,\phi')| \ .
\label{18l}
\end{equation}
Further, $|C(T,\phi,\phi')|$ is bounded from above by $\da^2$
according to (\ref{53}) and (\ref{18k}), yielding
\begin{equation} 
K \leq 
\da^2 \int_{\Gamma}d\phi \, \rho (E,\phi) \int_{\kam} d\phi'\, 
\rho (E,\phi') \ .
\label{18m}
\end{equation}
With (\ref{18d2}) and (\ref{18d1}) one readily sees that
the first integral on the right hand side of (\ref{18m}) 
is unity.
Exactly the same upper bound as in (\ref{18m})
is readily recovered for 
the modulus of the contribution from the third 
subdomain $\kam\times\sea$  to  (\ref{18j}).
Altogether we thus obtain the upper bound
\begin{equation}
\int dE \, p(E)\, \overline{\deltaA^2(E,t)}^\infty
\leq 2\,\da^2 \int dE\, p(E)\int_{\kam}
d\phi \, \rho(E,\phi) \ .
\label{18n}
\end{equation}
Taking into account (\ref{18d1}) and (\ref{18i1}) we finally can 
conclude that for any given $\epsilon>0$
\begin{equation}
\overline{\sigma^2(t)}^T\leq \epsilon + 2\,
\da^2 \int_{\kam} d\phi\, \rho_0(\phi)
\label{18o}
\end{equation}
for all sufficiently large $T$.

Next we define for any given $\dda>0$ and $T>0$ the
classical counterpart of (\ref{69}), namely
\begin{equation}
T_{\dda} := 
\big| \{0< t  <T\, : \, |\langle A\rangle_t-\langle A\rangle_{eq}|\geq\dda\}\big| \ ,
\label{69a}
\end{equation}
where $|M|$ denotes the size (Lebesgue measure) of the set $M$
and where $\langle A\rangle_t$ and $\langle A\rangle_{eq}$
are defined by (\ref{40a}) and (\ref{17b}), respectively.
Introducing $\epsilon':=\epsilon/\dda^2$ it 
follows as below (\ref{71}) 
that for any given $\epsilon'>0$ 
\begin{equation}
\frac{T_{\dda}}{T}\leq  \epsilon '
+ 2\, \left(\frac{\da}{\dda}\right)^2 
\int_{\kam} d\phi\, \rho_0(\phi)
\label{74a}
\end{equation}
for all sufficiently large $T$.

According to (\ref{69a}), the left hand side 
of (\ref{74a}) represents the fraction of all times
$t\in[0, T]$ for which there is an experimentally
resolvable difference between the true expectation 
value from (\ref{40a}) and the time-independent 
``equilibrium expectation value'' from (\ref{17b}).
In (\ref{74a}), the range-to-resolution ratio
of $\da/\dda$ is bounded by (\ref{9}),
while $\int_{\kam} d\phi\, \rho_0(\phi)$
is the initial population of the KAM-tori,
which is expected to become arbitrarily small
for a system with sufficiently many degrees 
of freedom $f$, see (\ref{31}).
In the generic case, this population 
$\int_{\kam} d\phi\, \rho_0(\phi)$
is furthermore expected to be non-zero
(see below (\ref{31})). Hence we may
choose $\epsilon'$ equal to the last 
term in (\ref{74a}) to arrive at
the classical counterpart of (\ref{74}), namely
\begin{equation}
\frac{T_{\dda}}{T}\leq 
4\, \left(\frac{\da}{\dda}\right)^2 
\int_{\kam} d\phi\, \rho_0(\phi)
\label{74b}
\end{equation}
for all sufficiently large $T$.

In either case, it follows with (\ref{69a})
-- analogously to (\ref{74}) --
that a system with sufficiently many degrees 
of freedom behaves as if it were in the 
``equilibrium state''
$\steady(\phi)$ for the overwhelming 
majority of times after initial transients 
have died out.

\subsection{Generalizations}
\label{s61}
In order to obtain (\ref{18j1}) we have
exploited the the mixing condition (\ref{1a}),
see also \cite{pen79}.
However, the same relation (\ref{18j1})
can readily be recovered under the 
alternative assumption that
\begin{equation}
\lim_{T\to\infty}
\overline{\left[\int_{\sea} \!\!\!\!\! d\phi\, \rho (E,\phi)\,\left\{A(\gamma_t(\phi))-\AAA(\phi)\right\}\right]^2}^T
= 0
\label{18j2}
\end{equation}
for all $\rho(E,\phi)$ of the form (\ref{18d1}) so that the 
phase space integral in (\ref{18j2}) is effectively confined 
to an energy surface with an arbitrary but fixed $E$
(see also \cite{f4}).
Property (\ref{18j2}) follows \cite{f6}
from the so called {\em weak mixing} condition \cite{hal49}
\begin{eqnarray}
\lim_{T\to\infty}
\overline{\left|\int_{\sea} \!\!\!\!\!  d\phi\, \rho (E,\phi)\,\left\{A(\gamma_t(\phi))-\AAA(\phi)\right\}\right|}^T
= 0 \ ,
\label{18j3}
\end{eqnarray}
which in turn quite obviously follows form the mixing 
condition (\ref{1a}) (which therefore is sometimes 
also referred to as {\em strong mixing} \cite{hal49}).

In summary, we thus found that either strong
mixing (\ref{1a}) or weak mixing (\ref{18j3})
(and taking for granted (\ref{9}) and (\ref{31})) 
is sufficient for equilibration 
in the sense outlined at the beginning of Sect. \ref{s3}.
However, it seems quite possible that a stronger
form of equilibration -- namely the asymptotic 
convergence of $\langle A\rangle_t$ for {\em all\,}
sufficiently large $t$ -- 
could follow under much weaker conditions
\cite{khi60,hob71,leb,bal71,rue91}.
First, in view of (\ref{40}) and (\ref{18a})
it is clear that if such a convergence 
does occur at all, then the limit will be given by (\ref{17b}).
Second, one readily sees that for any
(confining) system with a single degree of 
freedom ($f=1$), such a convergence indeed 
occurs in the generic case, i.e. for non-singular 
$\rho_0(\phi)$ 
(especially the energy
distribution $p(E)$ from (\ref{18d2}) 
must be non-singular)
and provided 
the frequency of the (analytically solvable)
periodic motion depends in a non-trivial way 
on the energy \cite{hob71}.
The same applies for a system consisting of 
several non interacting subsystems of 
this kind, see Chapter 3.4 in \cite{hob71}.
There seems no a priori reason that when the 
subsystems start to interact with each other, 
the long time limit of $\langle A\rangle_t$ 
should cease to exist.

In other words, the convergence of (\ref{18c})
to zero for $T\to\infty$ may well be true
under much weaker conditions than the
corresponding convergence on every single 
energy surface, as required by (\ref{18j2})
or by the weak mixing condition (\ref{18j3}).

\section{Thermalization in the Classical Case}
\label{s7}
As in Sect. \ref{s4}, the problem of thermalization
amounts -- in view of the conclusion below (\ref{74b}) --
to showing that the 
equilibrium
expectation value $\langle A\rangle_{eq}$
agrees, within the experimental resolution limit,
with the microcanonical expectation value 
\begin{equation}
\langle A\rangle_{mic}=\int d\phi \, \rho_{mic}(\phi)\, A(\phi) \ ,
\label{10}
\end{equation}
provided the energy 
distribution $p(E)$ from (\ref{18d2})
is peaked about its mean value $E_{mean}$ 
with a small but still experimentally 
realistic dispersion $\delta E$.
The microcanonical ensemble $\rho_{mic}(\phi)$
in (\ref{10}) is, as usual \cite{lldiu},
assumed to be equal to some normalization
constant if $H(\phi)$ is contained within the small 
energy interval $I$ from (\ref{103})
and zero otherwise.

In particular, all the unknown ``details'' 
(see Sect. \ref{s1}) of the usually 
out of equilibrium initial condition $\rho_0(\phi)$ 
are postulated to be irrelevant in (\ref{18d2}),
and likewise for the exact values of $E_{mic}$ 
and $\Delta E$ in (\ref{103})
(see also the discussion of property (P) in Sect.
\ref{s4}).
For the sake of simplicity the latter two quantities 
$E_{mic}$ and $\Delta E$ are thus
identified with their experimental counterparts
$E_{mean}$ and $\delta E$ from now on.

Intuitively, it is quite clear that the
energy distribution from (\ref{18d2})
remains unchanged as the system evolves in time.
Formally, this is readily confirmed by means of
(\ref{3}) yielding with (\ref{18a})
\begin{eqnarray}
p(E) & = &
\int d\phi\, \delta(H(\phi)-E) \rho(\phi,t)
\nonumber
\\ 
& = & \int d\phi\, \delta(H(\phi)-E) \steady(\phi) \ .
\label{60}
\end{eqnarray}
for arbitrary times $t$.

Next, we introduce yet another time-independent
probability density $\ssteady(\phi)$, 
which attributes the same probability
(\ref{60}) as the 
``true'' equilibrium ensemble $\steady(\phi)$
to any given ``energy surface'' and which is moreover
constant on any such surface,
\begin{equation}
\ssteady(\phi):=p(H(\phi))/\omega(H(\phi)) \ ,
\label{80}
\end{equation}
where the normalization $\omega (E)$ denotes the
``area'' of the energy surface,
\begin{equation}
\omega(E):=\int d\phi \  \delta(H(\phi)-E) \ ,
\label{70}
\end{equation}
and where (\ref{80}) is replaced by $\ssteady(\phi):=0$
in  case $\omega(H(\phi))=0$.
In other words, $\ssteady(\phi)$ most closely 
resemble the ``true'' $\rho_0(t)$ among all 
time-independent
densities which are constant on every energy surface. 
The corresponding expectation values are
denoted by (cf. eqs. (\ref{40}) and (\ref{17b}))
\begin{equation}
\langle A\rangle_{eq}^\ast:= \int  d\phi\, A(\phi)\, \ssteady(\phi) \ .
\label{40b}
\end{equation}

Our first main step in demonstrating thermalization
(see above (\ref{10})) is based on the remarkable,
completely general identity
\begin{equation}
\Abar-\langle A\rangle_{eq}^\ast
= \int d\phi\,  [\AAA (\phi)-\Aerg(H(\phi))] \, 
[\rho_0(\phi) - \ssteady(\phi)] \ ,
\label{26}
\end{equation}
where $\Aerg(H(\phi))$ is defined in (\ref{30}),
and whose derivation is provided in Appendix A.
Due to (\ref{30}), only $\phi\in \kam$ 
contribute to the integral in (\ref{26}), i.e.,
\begin{equation}
\langle A\rangle_{eq} - \langle A\rangle_{eq}^\ast
= \int_{\kam} \!\!\!\!\! d\phi\, [\AAA (\phi)-\Aerg(H(\phi))] \, 
[\rho_0(\phi) - \ssteady(\phi)] \ .
\label{28}
\end{equation}
Taking into account $\rho_0(\phi)\geq 0$, $\,\ssteady(\phi)\geq 0$,
and Eq. (\ref{53}) we thus obtain
\begin{equation}
|\Abar - \langle A\rangle_{eq}^\ast| \leq
\Delta_A \left[ \int_{\kam} \!\!\!\!\!  d\phi\, 
\rho_0(\phi) + \int_{\kam} \!\!\!\!\!  d\phi\, \ssteady(\phi)\right] \ .
\label{28a}
\end{equation}
The first integral becomes arbitrarily small 
for sufficiently large systems according to (\ref{31}).
The second integral is of the same type, except that
the original density $\rho_0(\phi)$ is now
uniformly redistributed within every 
energy surface, see below (\ref{70}).
Quite clearly, such a particularly ``tame''
density $\ssteady (\phi)$ instead of the ``true''
(possibly far from equilibrium)
initial density $\rho_0(\phi)$ is
actually also covered by (\ref{31}) 
as a special case.

In view of (\ref{9}) we thus can conclude from (\ref{28a})
that the the difference between $\Abar$
and $\langle A\rangle_{eq}^\ast$ becomes unmeasurably 
small for sufficiently many degrees of freedom.
Together with our previous result on equilibration
(see below (\ref{74a})) it follows that the system
behaves as if it were in the ``equilibrium state'' 
$\rho_{eq}^\ast(\phi)$ for the overwhelming majority 
of times $t$ after initial transients have died out.

Finally, we introduce 
the following average of 
an observable over an arbitrary but fixed
``energy surface'' (tacitly confining ourselves to 
$E$-values with $\omega(E)>0$),
\begin{equation}
A_\omega(E) := \int d\phi \, A(\phi)\,\delta(H(\phi)-E)/\omega(E)
\label{100}
\end{equation}
and -- analogously as in (\ref{18d2}), (\ref{60}) --
the microcanonical energy distribution
\begin{eqnarray}
p_{mic}(E) :=
\int d\phi\, \delta(H(\phi)-E) \rho_{mic}(\phi) \ .
\label{60a}
\end{eqnarray}
Combining these definitions with (\ref{10})-(\ref{80}) 
and (\ref{40b}) it follows that 
\begin{eqnarray}
\langle A\rangle_{eq}^\ast & = & 
\int dE\, p(E)\, A_\omega(E)
\label{105}
\\
\langle A\rangle_{mic} & = & 
\int dE\, p_{mic}(E)\, A_\omega(E) \ .
\label{105a}
\end{eqnarray}
These are the classical counterparts of eqs. 
(\ref{101}) and (\ref{102}).
Accordingly, the discussion below (\ref{102})
can be readily carried over to the present 
case.

In particular, both $p(E)$ and $p_{mic}(E)$ 
are assumed, as said below (\ref{10}), to
only take appreciable values within a small 
energy interval of width $\delta E$. 
Furthermore, the classical counterpart of
the ``eigenstate thermalization hypothesis'' 
from Sect. \ref{s4} now amounts to 
the condition that the variations 
of $A_\omega(E)$ from (\ref{100}) must
remain negligibly small (below the experimental
resolution limit $\dda$) when changing $E$ by 
less than the experimental energy resolution 
$\delta E$.
Intuitively, the quite general validity 
of this condition seems considerably more 
plausible than its quantum mechanical 
counterpart,
essentially amounting to some form of
uniform continuity of $A(\phi)$ along 
directions perpendicular to the 
energy surfaces.

We also remark that $\delta E$ is meant here as 
the remnant uncertainty after the experimentalist
has determined the energy of his system as 
accurately as possible. 
A violation of the above condition on 
$A_\omega(E)$ would imply that with the help of the 
observable $A(\phi)$ the energy uncertainty 
$\delta E$ could actually be reduced even 
further, in contradiction to the 
above meaning of 
$\delta E$. (This argument is not applicable 
in the quantum case since measuring $A$ would
change the state of the system.)

We finally note that the often considered 
limit $\delta E\to 0$ immediately implies
$\langle A\rangle_{eq}^\ast=A_\omega(E)=\langle A\rangle_{mic}$  
but is no way out: If $A_\omega(E)$ would
notably vary within the experimental 
energy uncertainty $\delta E$, this theoretical 
limit would be of little use to describe
the corresponding real experiment.
The fact that focusing on this limiting case 
is so often considered as sufficient actually
confirms once again that the above mentioned
small variations of $A_\omega(E)$ upon small
changes of $E$ have always been tacitly considered 
as a matter of course.

In any case, if and only if the difference between
(\ref{105}) and (\ref{105a}) is negligibly small
(below the experimental resolution limit)
thermalization can be inferred in the sense
that a system with sufficiently 
many degrees of freedom behaves as if it were
in the ``microcanonical state'' $\rho_{mic} (\phi)$
for the overwhelming majority of times 
after initial transients have died out.

\section{Summary and Discussion}
\label{s8}
Focusing on isolated macroscopic systems,
described either in terms of a
quantum mechanical or a classical 
model, our two key questions were:
In which sense and under which conditions 
does an initial ensemble (usually far from 
equilibrium and largely unknown in detail)
evolve towards a stationary long-time 
behavior (``equilibration'')?
In how far is this steady state in 
agreement with the microcanonical
ensemble predicted by equilibrium 
Statistical Mechanics (``thermalization'')?

Much like in equilibrium Statistical 
Mechanics itself, the concomitant
quantitative ``equilibration times'' are beyond 
the scope of our present approach.
Indeed, none of our calculations 
admits any meaningful conclusion
in this respect.
The reason is that any further quantification 
of the relaxation process inevitably would 
require considerably more detailed 
specifications of system (Hamiltonian),
initial condition (ensemble), and 
observables than in our present 
Sects. \ref{s2} and \ref{s5}.
In particular, the question why certain 
systems (glasses, Fermi-Pasta-Ulam 
model etc.) equilibrate so slowly that
they cannot be described in practice by
equilibrium Statistical Mechanics goes
beyond our present investigation.

In the Quantum case, equilibration 
was established in the sense 
that deviations from a time-independent
steady state become unmeasurably 
small for the overwhelming majority 
of times within any sufficiently 
large time interval.
In doing so, the main assumptions were:
finite range and resolution of observable;
no exceedingly large degeneracies of 
energy levels and energy gaps;
small populations of single energy 
levels by the initial ensemble.
Taking these prerequisites for granted
appears to be quite convincing to
model a real experiment whose details
are unavoidably unknown in practice.

In contrast, thermalization can so far be
shown only under additional, not
yet sufficiently well established 
assumptions, e.g. the eigenstate
thermalization hypothesis or 
Perez' argument, see Sect. \ref{s4},
whose validity and limitations
are presently at the focus of
numerous numerical and analytical
investigations
\cite{tas98,sre99,caz06,rig07,kol07,man07,cra08,bar08,rig08,spec,rou10,pal10,bri10,gog11,ike11,ban11,pol11,cam11,jac11,kas11,ji11,can11,kas12,rig12}.

In order to demonstrate equilibration in the 
same sense also in the classical case \cite{hob71},
we had to assume at least weak mixing 
within every energy surface 
(see (\ref{18j3})) and apart from
a subset of phase space consisting 
of the union of all KAM tori.
Further, we had to assume that
the initial ensemble populates this subset 
only with very small probability.
In support of both assumptions, quite
suggestive theoretical arguments and 
numerical evidence can be provided for 
systems with sufficiently many degrees 
of freedom (see Sect. \ref{s5}).
Regarding rigorous proofs,
the problems seems to be extremely difficult
in view of the quite limited progress
during several decades 
concerning general (generic) high 
dimensional Hamiltonian systems
\cite{skl93,for92,bun08,wig71,lic83,alt07}.

Once classical equilibration is established,
thermalization can be shown by means of
a relatively weak further assumption.
On the one hand, this assumption
may be viewed as the counterpart of the 
quantum mechanical eigenstate thermalization 
hypothesis, on the other hand one expects 
that it will be satisfied whenever the 
observable satisfies some rather mild 
continuity conditions (see Sect. \ref{s7})

As mentioned in Sect. \ref{s61},
it may well be that classical equilibration 
can be demonstrated under much weaker conditions
than those we have employed here \cite{khi60,hob71,leb,bal71,rue91}.
However, in order to subsequently demonstrate 
thermalization, eqs. (\ref{26}), (\ref{105}),
(\ref{105a}) suggest that at least ergodicity
(\ref{30}) and condition (\ref{31}) will 
still be required in the generic case, i.e.
unless (\ref{26}) happens to become
very small due to accidental 
cancellations.

In spite of the common opinion that
classical mechanics should follow from 
quantum mechanics,
our present results in the classical case 
cannot be deduced from those in the 
quantum case.
Generally speaking, the reason is that the
``classical limit'' $\hbar\to 0$ does not 
commute with the long time limit $T\to\infty$.
More specifically, the derivations 
of relations of the type (\ref{n2}) in 
\cite{rei08,lin09,lin10,rei10,sho11,sho12,rei12}
break down when taking the limit $\hbar\to 0$
while keeping the time $T$ finite.
Put differently, the energy spectrum must 
remain discrete and fixed while making $T$ 
``sufficiently large''.
For similar reasons, the quantum recurrence theorem
for mixed states (density operators)
mentioned at the beginning of Sect. \ref{s3}
does not survive in the classical limit
(the Poincar\'e recurrence theorem 
only concerns pure states).
The situation is comparable to the relation 
between quantum chaos and classical 
Hamiltonian chaos:
In the quantum case the main tools are level
statistics and random matrix theory and 
the obtained results do not allow one 
to recover the key features of chaos in 
the classical limit, namely KAM-theory and 
its implications regarding Lyapunov 
exponents and ergodicity.
Accordingly, completely different approaches
in the quantum and the classical case
are also required and provided in the 
present work.

In particular, while our present 
quantum mechanical approach includes 
pure states as special cases 
(see below Eq. (\ref{7})),
the classical counterpart breaks down for 
pure states (see below Eq. (\ref{0})).
In other words, while our present classical
approach deals with ``instantaneous''
measurements on non-trivial statistical 
ensembles (mixed states), classical 
ergodic theory is mainly concerned with
time-averaged properties of pure states
(see also Sec. \ref{s1}).

A hierarchy of increasingly strong ``stochasticity 
properties'' are commonly used to characterize 
the degree of ``chaoticity'' of a classical 
Hamiltonian system, the weakest being ergodicity, 
followed by weak mixing, strong mixing, K-systems, 
Anosov-systems etc. \cite{wig71,skl93}.
Our present classical approach employs the first
(weakest) three categories.
The corresponding notions in the quantum case 
are not very clearly defined.
Comparing our present findings in the
quantum and the classical cases naturally
suggests certain ``analogies'' or ``correspondences''
along these lines, but they do not seem 
to provide any additional physical insight.

\begin{center}
\vspace{0mm}
---------------------------
\vspace{0mm}
\end{center}
Specials thanks is due to Sheldon Goldstein 
for extremely insightful criticisms and hints.
This work was supported by 
Deutsche Forschungsgemeinschaft 
under grant RE1344/7-1.

\section*{Appendix A}
This Appendix provides the derivation
of eq. (\ref{26}).

By means of (\ref{18d2}) and (\ref{80}) we can 
rewrite (\ref{40b}) as
\begin{equation}
\langle A\rangle_{eq}^\ast
= \int d\phi\, A(\phi) \int d\phi' \, \rho_0(\phi')
\, \frac{\delta(H(\phi')-H(\phi))}{\omega(H(\phi))} \ .
\label{100a}
\end{equation}
By exploiting the definition (\ref{100})
we furthermore can conclude that
\begin{equation}
\langle A\rangle_{eq}^\ast
=  \int d\phi' \, \rho_0(\phi') \, A_{\omega}(H(\phi')) \ .
\label{16a}
\end{equation}
Eqs. (\ref{18b}) and (\ref{16a}) finally yield
\begin{equation}
\langle A\rangle_{eq}-\langle A\rangle_{eq}^\ast = 
\int d\phi\, [\AAA(\phi)- A_\omega(H(\phi))]\, \rho_0(\phi) \ .
\label{20}
\end{equation}

Next we can infer from (\ref{3}) that $A(\phi)$ on the right
hand side of (\ref{100}) can be replaced by $A(\gamma_t(\phi))$
for any $t$, and thus it can also be replaced by 
the time average from (\ref{1b}), i.e.
\begin{equation}
A_{\omega}(E):=\int d\phi \, \AAA (\phi) \, 
\frac{\delta(H(\phi)-E)}{\omega(E)} \ .
\label{18z}
\end{equation}
Further, recalling the definition of $\omega(E)$ from (\ref{70}), 
one readily verifies that
\begin{equation}
\Aerg(E)=\int d\phi \, \Aerg(H(\phi))\,\frac{\delta(H(\phi)-E)}{\omega(E)} \ .
\label{21}
\end{equation}
In combination with (\ref{18z}) we thus obtain
\begin{eqnarray}
A_{\omega}(E) 
& = & \Aerg(E)+\int d\phi \, D(\phi) \, \frac{\delta(H(\phi)-E)}{\omega(E)}
\label{22}
\\
D(\phi) & := & \AAA (\phi)-\Aerg(H(\phi)) \ .
\label{22b}
\end{eqnarray}
Introducing (\ref{22}) and (\ref{22b}) into (\ref{20}) yields
\begin{eqnarray}
& & \Abar-\langle A\rangle_{eq}^\ast = \int d\phi\, D(\phi)\, \rho_0(\phi) - J
\label{23}
\\
& & J  :=  \int d\phi \int d\phi' \, 
D(\phi') \, \frac{\delta(H(\phi')-H(\phi))}{\omega(H(\phi))}\, \rho_0(\phi)
\nonumber
\\
& & =
\int d\phi' \, D(\phi') \, 
\frac{\int d\phi \, \delta(H(\phi')-H(\phi)) \, \rho_0(\phi)}{\omega(H(\phi'))} 
\label{24}
\end{eqnarray}
The last integral over $\phi$ equals $p(H(\phi'))$ according 
to (\ref{18d2}) and after
division by $\omega(H(\phi'))$ we recover $\ssteady(\phi')$ from (\ref{80}).
Omitting the primes, (\ref{24}) thus takes the form
\begin{eqnarray}
J = \int  d\phi \, D(\phi) \, \ssteady(\phi) 
\label{25}
\end{eqnarray}
With (\ref{22b}) and (\ref{23}) we finally
recover (\ref{26}).


\begin{thebibliography}{99}

\bibitem{skl93}
L. Sklar, {\em Physics and Chance} (Cambridge University Press, 1993).

\bibitem{pen79}
O. Penrose, Rep. Prog. Phys. {\bf 42}, 129 (1979).

\bibitem{gem09}
J. Gemmer, M. Michel, and G. Mahler, {\em Quantum Thermodynamics}
(2nd edition, Springer, Berlin, Heildelberg, 2009).

\bibitem{wig71}
A. S. Wightman, pp 1-32 in 
{\em Statistical Mechanics at the turn of the Decade},
edited by E. G. D. Cohen (Marcel Dekker Inc., New York, 1971).

\bibitem{wal}
D. Wallace,
{\em Implications of quantum theory in the foundations 
of statistical mechanics},
http://philsci-archive.pitt.edu/410 (2001). 

\bibitem{far64}
I. E. Farquhar, {\em Ergodic Theory in Statistical Mechanics} (Interscience, NY, 1961).

\bibitem{jan69}
R. Jancel, {\em Foundations of Classical and Quantum 
Statistical Mechanics} (Pergamon, London, 1969).

\bibitem{spec}
Focus Issue {\em Dynamics and thermalization in 
isolated quantum many-body systems},
edited by M. A. Cazalilla and M. Rigol
New J. Phys. {\bf 12}, (2010).

\bibitem{pol11}
A. Polkovnikov, K. Sengupta, A. Silva, and M. Vengalattore,
Rev. Mod. Phys. {\bf 83}, 863 (2011).

\bibitem{yuk11}
V. I. Yukalnov, Laser Phys. Lett. {\bf 8}, 485 (2011).

\bibitem{rei08}
P. Reimann, Phys. Rev. Lett. {\bf 101}, 190403  (2008).

\bibitem{lin09}
N. Linden, S. Popescu, A. J. Short, and A. Winter, 
Phys. Rev.  E {\bf 79}, 061103 (2009).

\bibitem{lin10}
N. Linden, S. Popescu, A. J. Short, and A. Winter, 
New J. Phys. {\bf 12}, 055021 (2010).

\bibitem{rei10}
P. Reimann, New J. Phys. {\bf 12}, 055027 (2010).

\bibitem{sho11}
A. J. Short, New J. Phys. {\bf 13}, 053009  (2011).

\bibitem{sho12}
A. J. Short and T. C. Farrelly, New J. Phys. {\bf 14}, 013063 (2012).

\bibitem{rei12}
P. Reimann and M. Kastner, New J. Phys. {\bf 14}, 043020 (2012).

\bibitem{tas98}
H. Tasaki, Phys. Rev. Lett. {\bf 80}, 1373 (1998).

\bibitem{sre99}
M. Srednicki, J. Phys. A: Math. Gen. {\bf 32}, 1163 (1999).

\bibitem{caz06}
M. A. Cazalilla, Phys. Rev. Lett. {\bf 97}, 156403 (2006).

\bibitem{rig07}
M. Rigol, V. Dunjko, V. Yurovsky, and M. Olshanii,
Phys. Rev. Lett. {\bf 98}, 050405 (2007).

\bibitem{kol07}
C. Kollath, A. M. L\"auchli, and E. Altman, 

Phys. Rev. Lett. {\bf 98}, 180601 (2007).

\bibitem{man07}
S. R. Manmana, S. Wessel, R. M. Noack, and A. Muramatsu, 
Phys. Rev. Lett. {\bf 98}, 210405 (2007).

\bibitem{cra08}
M. Cramer, C. M. Dawson, J. Eisert, and T. J. Osborne, 
Phys. Rev. Lett. {\bf 100}, 030602 (2008).

\bibitem{bar08}
T. Barthel and U. Schollw\"ock,
Phys. Rev. Lett. {\bf 100}, 100601 (2008).

\bibitem{rig08}
M. Rigol, V. Dunjko, and M. Olshanii, 
Nature {\bf 452}, 854 (2008).

\bibitem{rou10}
G. Roux, Phys. Rev.  A {\bf 81}, 053604 (2010).

\bibitem{pal10}
A. Pal and D. A. Huse, Phys. Rev.  B {\bf 82}, 174411 (2010).

\bibitem{bri10}
G. Biroli, C. Kollath, and A. M. L\"auchli, Phys. Rev. Lett. {\bf 105}, 250401 (2010).

\bibitem{gog11}
C. Gogolin, M. P. M\"uller, and J. Eisert,
Phys. Rev. Lett. {\bf 106}, 040401 (2011).

\bibitem{ike11}
T. N. Ikeda, Y. Watanabe, and M. Ueda,
Phys. Rev.  E {\bf 84}, 021130 (2011).

\bibitem{ban11}
M. C. Ba\~{n}uls, J. I. Cirac, and M. B. Hastings, 
Phys. Rev. Lett. {\bf 106}, 050405 (2011).

\bibitem{cam11}
L. Campos Venuti, N. T. Jacobson, S. Santra, and P. Zanardi,
Phys. Rev. Lett. {\bf 107}, 010403 (2011).

\bibitem{jac11}
N. T. Jacobson, L. C. Venuti, and P. Zanardi,
Phys. Rev.  A {\bf 84}, 022115 (2011).

\bibitem{kas11}
M. Kastner, Phys. Rev. Lett. {\bf 106}, 130601 (2011).

\bibitem{ji11}
K. Ji and B. V. Fine, Phys. Rev. Lett. {\bf 107}, 050401 (2011).

\bibitem{can11}
E. Canovi, D. Rossini, R. Fazio, G. E. Santoro, and A. Silva,
Phys. Rev.  B {\bf 83}, 094431 (2011).

\bibitem{kas12}
M. Kastner, Cent. Eur. J. Phys. {\bf 10}, 637 (2012).

\bibitem{rig12}
M. Rigol and M. Srednicki, Phys. Rev. Lett. {\bf 108}, 110601 (2012).

\bibitem{khi60}
A. Y. Khinchin, {\em Mathematical Foundations of Quantum Statistics}
(Graylock Press, New York, 1960).

\bibitem{leb}
J. L. Lebowitz,
{\em From Time-symmetric Microscopic Dynamics to 
Time-asymmetric Macroscopic Behavior: An Overview},
arXiv:0709.0724v1 (2007);
S. Goldstein, {\em Boltzmann's Approach to Statistical Mechanics},
arXiv:cond-mat/0105242v1 (2001);
S. Goldstein, J. L. Lebowitz, C. Mastrodonato, 
R. Tumulka, and N. Zangh\`{\i}, Phys. Rev.  E {\bf 81}, 011109 (2010);
S. Goldstein, J. L. Lebowitz, R. Tumulka, and N. Zangh\`{\i}, 
Eur. J. Phys. H {\bf 35}, 173 (2010);
S. Goldstein, J. L. Lebowitz, C. Mastrodonato,
R. Tumulka, and N. Zangh\`{\i},
{\em Proc. R. Soc. London Ser. A} {\bf 466}, 3203 (2010);
H. Tasaki, {\em The approach to thermal equilibrium and 
``thermodynamic nor\-mal\-i\-ty''}, 
arXiv:1003.5424v4 (2010).

\bibitem{hob71}
A. Hobson, {\em Concepts in Statistical Mechanics} 
(Gordon and Breach, New York, 1971).

\bibitem{mar78}
L. Markus and K. R. Mayer, Mem. Am. Math. Soc. {\bf 144}, 1 (1978).

\bibitem{bun08}
L. A. Bunimovich, Nonlinearity {\bf 21}, T13 (2008).

\bibitem{realobs2}
E. P. Wigner, Am. J. Phys. {\bf 31}, 6 (1963).

\bibitem{realobs3}
N. G. van Kampen, chapter XVII.7 in
{\em Stochastic Processes in Physics and Chemistry}
(Elsevier, Amsterdam, 1992).

\bibitem{realobs4}
A. Sugita, Nonlinear Phenom. Complex Syst. {\bf 10}, 192 (2007).

\bibitem{realobs5}
O. Penrose, chapter 1 in {\em Foundations of Statistical Mechanics} 
(Pergamon, Oxford, 1970).

\bibitem{lof}
A. Martin-L\"of, {\em Statistical Mechanics and the 
Foundations of Thermodynamics}
(Lecture Notes in Physics Vol. 101, Berlin, Springer, 1979).

\bibitem{geo95}
H.-O. Georgii, J. Stat. Phys. {\bf 80}, 1341 (1995).

\bibitem{pop06} 
S. Popescu, A. J. Short, and A. Winter, Nature Phys. {\bf 2}, 754 (2006).

\bibitem{rei08a}
P. Reimann, J. Stat. Phys. {\bf 132}, 921 (2008).

\bibitem{lldiu}
L. D. Landau and E. M. Lifshitz, {\em Statistical Physics}
(Pergamon, Oxford 1980);
B. Diu, C. Guthmann, D. Lederer, and B. Roulet, 
{\em Elements de Physique Statistique} 
(Hermann, Paris, 1996).

\bibitem{lud58}
G. Ludwig,  Z. Phys. {\bf 150}, 346 (1958); {\bf 152}, 98 (1958).

\bibitem{boc59}
P. Bocchieri and A. Loinger, Phys. Rev. {\bf 114}, 948 (1959).

\bibitem{per84}
A. Peres, Phys. Rev. A {\bf 30}, 504 (1984).

\bibitem{deu91}
J. M. Deutsch, Phys. Rev. A {\bf 43}, 2046 (1991).

\bibitem{sre96}
M. Srednicki, J. Phys. A: Math. Gen {\bf 29}, L75 (1996);
{\bf 32} 1163 (1996).

\bibitem{f1}
Note that no such degeneracies 
are actually encountered in the generic case
\cite{per84,sre99,tas98,gol06,rei08,lin09,rei10},
i.e. in the absence of special reasons like
additional conserved 
quantities (e.g. due to (perfect) symmetries or 
when the system consists of non-interacting 
subsystems) or fine-tuning of parameters 
(``accidental degeneracies'').

\bibitem{gol06}
S. Goldstein, J. L. Lebowitz, J. L. Tumulka, and N. Zangh\`{\i},
J. Stat. Phys. {\bf 125}, 1197 (2006).

\bibitem{f1a}
In particular, $T$ must obviously be much larger
than the relaxation time in case
of a far-from-equilibrium initial 
condition $\rho(0)$.

\bibitem{fei86}
M. Feingold and A. Peres, Phys. Rev. A {\bf 34}, 591 (1986).

\bibitem{sre94} 
M. Srednicki, Phys. Rev. E {\bf 50}, 888 (1994);
cond-mat/9410046

\bibitem{jen85}
R. V. Jensen and R. Shankar, Phys. Rev. Lett. {\bf 54}, 1879 (1985).

\bibitem{fei84}
M. Feingold, N. Moiseyev, and A. Peres, Phys. Rev. A {\bf 30}, 509 (1984).

\bibitem{wei92}
S. Weigert, Physica D {\bf 56}, 107 (1992);
B. Sutherland, Ch. 2.1 in {\em Beautiful Models} 
(World Scientific, Singapore, 2004);
A. Enciso  and D. Peralta-Salas, 
Theor. Math. Phys. {\bf 148}, 1086 (2006).

\bibitem{for92}
J. Ford, Phys. Rep. {\bf 213}, 271 (1992).

\bibitem{lic83}
A. L. Lichtenberg and M. A. Liebermann, Sect. 6.5 in
{\em Regular and Chaotic Motion}
(Springer, Berlin, 1983).

\bibitem{har40}
P. Hartman and A. Wintner, Amer. J. Math. {\bf 62}, 646 (1940)

\bibitem{hal49}
P. R. Halmos, Bulletin of the American Mathematical Society 
{\bf 55}, 1015 (1949).

\bibitem{f2} 
The possibility to replace 
(\ref{1a}) by substantially weaker conditions will 
be discussed in Sect. \ref{s61}.

\bibitem{alt07}
C. Froeschl\'e and J.-P. Scheidecker, Phys. Rev. A {\bf 12},
2137 (1975);
M. Falcioni, U. Marconi, and A. Vulpiani, Phys. Rev. A {\bf 44}, 2263 (1991);
E. G. Altmann and H. Kantz, EPL {\bf 78}, 10008 (2007).

\bibitem{bal71}
R. Balescu, {\em Equilibrium and Nonequilibrium Statistical Mechanics}
(Wiley, New York, 1971).

\bibitem{f3} 
Our present notion of 
equilibration is thus very similar to the one 
by Hobson (see Chapter 5.2 in \cite{hob71}),
which is based on the concept of 
weak convergence.
Its close relation to the notion of ``coarse 
graining'' is discussed e.g. in \cite{pen79}.

\bibitem{f4} 
The $\delta$-function 
in (\ref{18d1}) is a convenient
formal way to effectively project the Euclidean Lebesgue 
measure of the phase-space on the energy surfaces in the 
usual way, see e.g. p. 22-23 in \cite{lev34} and 
p. 1945-1946 in \cite{pen79}.

\bibitem{lev34}
T. Levi-Civita, Journal of Mathematics and Physics {\bf 34}, 18 (1934).

\bibitem{f5}
Similarly, also the convergence of 
$\overline{\sigma^2(t)}^T$ for $T\to\infty$ 
can be deduced from (\ref{18c}).
In both cases, a rigorous justification of 
commuting integrations with the $T\to\infty$ 
limit is, however, not obvious at all.
It is closely related with the question
whether or not the characteristic convergence
time of those quantities will be comparable
to the notoriously large 
``ergodicity time scales'' governing the 
convergence in (\ref{1b}) and (\ref{1a}).
The analogous problem in the quantum case is 
treated e.g. in Sect. 4.3 of \cite{rei12}.

\bibitem{f6}
Abbreviating
the integral appearing in (\ref{18j2}) as $h(t)$,
it follows with (\ref{53}) that $|h(t)|$ 
can be bounded from above by some finite 
constant $c$ for all $t$.
Hence $\overline{[h(t)]^2}^T\leq c\, \overline{|h(t)|}^T$.
As a consequence, (\ref{18j3}) implies (\ref{18j2}).

\bibitem{rue91}
D. Ruelle, {\em Chance and Chaos} 
(Princeton University Press, Princeton NJ, 1991);
L. Bunimovich, Chaos {\bf 13}, 903 (2003).

\end{thebibliography}
\end{document}